\documentclass[acmsmall,manuscript,screen]{acmart}

\AtBeginDocument{%
  }

\setcopyright{acmlicensed}
\copyrightyear{2025}
\acmYear{2025}
\acmDOI{XXXXXXX.XXXXXXX}

\acmConference[Conference acronym 'XX]{Make sure to enter the correct
  conference title from your rights confirmation emai}{June 03--05,
  2018}{Woodstock, NY}
\acmISBN{978-1-4503-XXXX-X/18/06}

\usepackage{tikz}
\usepackage{amsmath}

\usepackage{graphicx}
\usepackage{subcaption}
\usepackage{csquotes}
\usepackage[inline]{enumitem}
\usepackage{booktabs}
\usepackage{algorithm2e}
\usepackage{multirow}

\newcommand\litem[1]{\item{\itshape #1}}

\usepackage{soul}

\newcommand{\rev}[1]{#1}
\newcommand{\del}[1]{}

\begin{document}

\title{Libertas: Privacy-Preserving Collaborative Computation for Decentralised Personal Data Stores}


\author{Rui Zhao}
\affiliation{%
  \institution{University of Oxford}
  \country{United Kingdom}}
\email{rui.zhao@cs.ox.ac.uk}
\orcid{0000-0003-2993-2023}

\author{Naman Goel}
\affiliation{%
  \institution{University of Oxford}
  \country{United Kingdom}}
\email{naman.goel@cs.ox.ac.uk}
\orcid{0000-0002-5106-5889}

\author{Nitin Agrawal}
\authornote{This author contributed to this work while being affliated with the University of Oxford}
\affiliation{%
  \institution{University of Oxford}
  \country{United Kingdom}}
\email{nitin.cic@gmail.com}

\author{Jun Zhao}
\affiliation{%
  \institution{University of Oxford}
  \country{United Kingdom}}
\email{jun.zhao@cs.ox.ac.uk}

\author{Jake Stein}
\affiliation{%
  \institution{University of Oxford}
  \country{United Kingdom}}
\email{jake.stein@cs.ox.ac.uk}

\author{Wael Albayaydh}
\affiliation{%
  \institution{University of Oxford}
  \country{United Kingdom}}
\email{wael.albayaydh@cs.ox.ac.uk}

\author{Ruben Verborgh}
\affiliation{%
  \institution{Ghent University}
  \country{Belgium}}
\email{Ruben.Verborgh@UGent.be}

\author{Reuben Binns}
\affiliation{%
  \institution{University of Oxford}
  \country{United Kingdom}}
\email{reuben.binns@cs.ox.ac.uk}

\author{Tim Berners-Lee}
\affiliation{%
  \institution{University of Oxford}
  \country{United Kingdom}}
\email{tim.berners-lee@cs.ox.ac.uk}

\author{Nigel Shadbolt}
\affiliation{%
  \institution{University of Oxford}
  \country{United Kingdom}}
\email{nigel.shadbolt@cs.ox.ac.uk}

\renewcommand{\shortauthors}{Zhao et al.}

\begin{abstract}
Data and data processing have become an indispensable aspect for our society. Insights drawn from collective data make invaluable contribution to scien\rev{tific}, \del{medical, cultural,}societal and communal research and business. But there are increasing worries about privacy issues and data misuse. This has prompted the emergence of decentralised personal data stores (PDS) like Solid that provide individuals more control over their personal data. However, existing PDS frameworks face challenges in ensuring data privacy when performing collective computations that combine data from multiple users. While Secure Multi-Party Computation (MPC) offers input secrecy protection during collective computation without relying on any single party, issues emerge when directly applying MPC in the context of PDS, particularly due to key factors like autonomy and decentralisation.
In this work, we discuss the essence of this issue, identify a potential solution, and introduce a modular system architecture, Libertas, to integrate MPC with PDS like Solid, without requiring protocol-level changes. We introduce a paradigm shift from \rev{an `}omniscien\rev{t' view} to individual-based\rev{,} user-centric \rev{view of} trust \rev{and security}, and discuss the threat model of Libertas.
Two realistic use cases for collaborative data processing are used for evaluation, both for technical feasibility and empirical benchmark, highlighting its effectiveness in empowering gig workers and generating differentially private synthetic data.
The results of our experiments underscore Libertas' linear scalability and provide valuable insights into compute optimisations, thereby advancing the state-of-the-art in privacy-preserving data processing practices. By offering practical solutions for maintaining both individual autonomy and privacy in collaborative data processing environments, Libertas contributes significantly to the ongoing discourse on privacy protection in data-driven decision-making contexts.
\end{abstract}




\maketitle

\section{Introduction}
\label{sec:intro}

Data is a valuable resource for science and business, \rev{as well as} for society and individuals. There is a growing imperative to leverage collective data for societal benefit\rev{s}. Examples include enhancing pandemic responses by aggregating health records~\cite{troncoso_decentralized_2020}, addressing climate change through energy consumption data exchange~\cite{baeck_using_2020}, and even ameliorating working conditions by pooling telemetry data among otherwise isolated gig workers \cite{zhang_algorithmic_2022,stein_you_2023}.

Previous CSCW \rev{research} has extensively discussed the benefits of user-centric or user-led approaches to collaboration \rev{in} extracting collective insights \cite{calacci_bargaining_2022,lam_enduser_2022,gupta_instagram_2022,zieglmeier_rethinking_2023}, which provided evidence and suggested the necessity of helping users regain control over their data and achieving better data autonomy.
However, most often, they still make use of a trusted central party (the system) for data storage, data management and computation.
This can lead to issues and concerns similar to existing centralised platforms, such as barrier of account registration \cite{hill_hidden_2021,jay_sign_2016,malheiros_sign-up_2013}, lack of data portability \cite{jamieson_escaping_2023,shipman_ownership_2020}, and trust and abuse concerns \cite{afriat_this_2021}.
A truly user-centric design should eliminate such dependencies and \rev{their} associated concerns, fostering better user adoption and \rev{greater} flexibility \rev{in} engaging with or withdrawing from \rev{both} new \rev{and existing} collective computational tasks.

Two recent developments that have gained increasing attention could form potential foundations for addressing these issues: decentralised Personal Data Stores and privacy-preserving computation.
Personal Data Stores (PDS) \cite{2014openpds,2016databox,2016solid} are considered a promising decentralised approach to addressing the privacy and autonomy concerns faced by individuals due to centralised platforms \cite{2021konradgdpr,zuboff2019age}. They promote a paradigm shift from \textit{platform-centred} architectures to \textit{user-centric} architectures, where users store\del{s} their data in a trusted location (the PDS), and applications or services \del{make} request\del{s to} access to \rev{such} data from the PDS\rev{,} \del{(}rather than storing and managing \rev{data} on their own servers\del{)}. 
However, this model still \rev{presents} privacy challenges when collaborative computation is \rev{required} on data distributed across multiple PDSs (see Section \ref{sec:background:pds}).
Thus, algorithmic mechanisms for privacy-preserving computation offer a promising choice to address these challenges with mathematical guarantees.
In particular, Secure Multi-Party Computation (MPC) \cite{yao_protocols_1982,lindell_secure_2020} enables the secure evaluation of functions \rev{over} multiple parties' data without revealing \rev{their} inputs. This can be further complemented by Differential Privacy (DP) \cite{dwork_differential_2008}\rev{,} which aims to mitigate concerns about inferring input\rev{s} from computational outputs.
However, discussions in the relevant fields have \rev{so far} mainly focused on \del{the} potential benefits, such as data quality and performance\rev{. A}s we will see later (Section \ref{sec:background_privacy} and Section \ref{sec:mpc-challenges-with-pds}), challenges \rev{remain} in a user-autonomous setting.

This paper seeks to answer a natural yet \rev{overlooked} question: \rev{I}s it feasible to perform collaborative computation in a decentralised PDS context while respecting ethical aspects of privacy?
In particular, can we go beyond the data secrecy / confidentiality aspect of privacy, and also include users' \rev{varying preferences for control, autonomy and trust} as dynamically customisable factors \rev{in} a system?

We identify several core ethical and functional requirements that such a solution should address, as summarised in Table \ref{tab:req_ppc}.
As we will discuss in Section \ref{sec:background:ethical-concept}, we \rev{incorporate} additional ethical \rev{dimensions} of privacy into technical systems: the traditional CS aspect of confidentiality is referred to as the (input) \textbf{data privacy (R1)} requirement, which regulates the computational mechanism to not unexpectedly leak data; \emph{user's will and control} is referred to as \textbf{(user) autonomy (R2)}, which expresses additional constraints that the technical solution should satisfy when performing computation\rev{s}; we also highlight the need for \textbf{user-centric trust (R3)}, that the system should respect and not coerce different users' heterogeneous trust preferences. In addition, two functional requirements \rev{have} also \rev{been} identified: the system should possess good \textbf{scalability (R4)} with \rev{an} increasing number of participants, to be deployable in \rev{the} real world; likewise, the system should be generic (\del{the }\textbf{generality \del{[R5]}} requirement \rev{\textbf{R5}}) to support various (if not all) types of computation, rather than \rev{being} specialised to a limited number of computational jobs. Of course, there are other aspects that a system can support, but we consider the said requirements as the key requirements for a generic system (see also Section~\ref{sec:background:ethical-concept} for further critical discussion\rev{s} about requirements). We demonstrate in th\rev{is} paper that our proposed system, Libertas, can address the five requirements, and also support additional aspects, such as \textit{output privacy}, in addition\del{al} to the \textit{input privacy}, enabling \rev{further uses}.

\begin{table*}
    \centering
    \caption{Identified crucial ethical and functional requirements for a general collective privacy-preserving computation system in decentralised context\rev{s}.}
    \label{tab:req_ppc}
    \begin{tabular}{ll}
        \toprule
        \textbf{Requirement} & \textbf{Explanation} \\
        \midrule
        R1: Data Privacy & Keeps (input) data confidential for the computation, not revealing to third parties \\
        R2: User Autonomy & Gives meaningful control to the data providers/owners over permitted usage\rev{s} of their data\\
        R3: User-centric Trust & \begin{tabular}{@{}l@{}}Respects heterogeneous trust preferences by different users, and modifies system \\ behaviour\rev{s} accordingly\end{tabular} \\
        R4: Scalability & Scales well with the number of data providers \\
        R5: Generality & Supports various types of computations \\
        \bottomrule
    \end{tabular}
\end{table*}

In the \rev{remainder} of this paper, we identify the shortcomings of current practices (Section \ref{sec:background}) and the challenges \rev{of} combining MPC with PDS (Section \ref{sec:mpc-challenges-with-pds}). These challenges stem from the Autonomy (R2) and Minimum Trust (R3) requirements\rev{,} and the involvement of numerous autonomous data providers not typically encountered in centralised settings. We introduce Libertas (Section \ref{sec:solid_mpc}), a pioneering modular architecture that addresses \textbf{how} to combine MPC with PDS to support \rev{collaborative} privacy-preserving computation in decentralised contexts, without requiring changes to underlying protocols. We demonstrate our implementation based on Solid \cite{2016solid}, and discuss its adaptability to other PDS systems.
\del{The proposed architecture utilises the delegated-decentralised MPC model, offering superior scalability (linearly with the number of data providers) compared to the direct-decentralised model, as validated through benchmarking (Section \ref{sec:benchmark:mpc-models})}. Individual-based\rev{,} user-centric trust is a unique property of Libertas, where trust preference\rev{s are} specified by users individually\rev{. W}e discuss its implications in the threat model \rev{section} (Section \ref{sec:solid_mpc:threats}). \rev{Libertas utilises the delegated-decentralised MPC model, offering superior scalability (linearly with the number of data providers) compared to the direct-decentralised model, as validated through benchmarking (Section \ref{sec:benchmark:mpc-models}).} We further evaluate \del{the }Libertas \del{architecture }with two realistic example use cases (Section \ref{sec:evaluation:solid-mpc}): 1) gig workers performing collective computation on their earnings data\rev{;} and 2) synthetic differentially private data generation for privacy-enhancing public data release (\del{for }e.g. census data). These evaluations confirm the technical feasibility and scalability of our approach, showcasing its broad applicability and potential.

\paragraph*{Contributions:} This paper makes both theoretical and practical contributions, as summarised below:
\begin{enumerate}
    \item It highlights a distinct challenge and gap in employing Secure Multi-Party Computation (MPC) in decentralised user-autonomous contexts such as Personal Data Stores (PDS);
    \item It proposes a novel technical solution, Libertas\footnote{Our prototype implementation is available at \url{https://github.com/OxfordHCC/libertas}.}
    , to address the challenge, \rev{and} highlights the necessity and implications of individual-based\rev{,} user-centric \rev{security and} trust, in contrast to the \rev{`}omniscien\rev{t'} \rev{view} in existing literature;
    \item Through two case studies of collaborative computations \rev{--} gig worker empowerment and synthetic data generation \rev{--} it evaluates the proposed system as follows:
    \begin{itemize}
        \item Theoretically, \rev{it} demonstrate\rev{s} how Libertas can contribute to CSCW for achieving collective benefits \rev{while} strengthening individual privacy and autonomy simultaneously. Our discussion in the gig worker case study also considers how real-life power dynamics can \rev{establish} the required trust relationships in Libertas to foster adoption;
        \item Practically, \rev{it} validate\rev{s} the implementation, demonstrate\rev{s} scalability patterns, and identif\rev{ies} practical insights.
    \end{itemize}
\end{enumerate}

\section{Background and Related Work}
\label{sec:background}

As our work aims to bridge a gap between socio-technical discussions and technical solutions \rev{regarding} privacy, and \rev{because} there are some differences in \rev{terminology} used in the fields, this section first presents a general discussion \rev{of these differences}, and then \rev{introduces} the \del{relevant} technologies \rev{relevant} to our research.

\subsection{Privacy and Trust}
\label{sec:background:ethical-concept}

\paragraph{Different aspects of privacy}

The concept of privacy ha\rev{s} been \rev{widely} researched\del{ widely}, and interdisciplinary scholars have noted the difficulty of precisely defining privacy \cite{smith_information_2011,ohara_seven_2016,regan_legislating_1995, nissenbaum_contextual_2011,decew_pursuit_1997}. In socio-technical systems, (information) privacy practices often include informational self-determination, the autonomy to control access to the information about self, and also contextual norms~\cite{nissenbaum_contextual_2011, decew_pursuit_1997, smith_information_2011, albayaydh_examining_2023}.

Computer science research often considers security and privacy simultaneously, focusing on a narrower aspect of (information) privacy: information (data) confidentiality. Examples include cryptography \cite{diffie_new_1976}, multi-party computation \cite{lindell_secure_2020}\rev{,} and differential privacy \cite{dwork_differential_2008}, where \del{the term} \rev{`}privacy\rev{'} refers to data (or other types of information) being kept secret (accessible only to the ``owner" or other prescribed parties as discussed in the respective technology). While not sufficient \rev{on their own}, these privacy-enhancing technologies (PETs) can provide \rev{individuals with} a complementary means \del{to individuals} \rev{of} protect\rev{ing} their personal information, while also \rev{enabling them} to use that information for individual and collective benefit\del{s}.



In this work, we attempt to bridge the gap between socio-technical discussions of privacy (specifically, autonomy or self-control) and practices in computing (specifically, confidentiality of data used in computation), as summarised in Table \ref{tab:req_ppc}. Note that we distinguish autonomy from the other privacy aspects conventionally discussed in computer science literature\rev{,} us\rev{ing} a different term to \rev{highlight} this gap; this is not to imply that autonomy is an entirely \rev{separate} concept from privacy in \rev{the} broad sense.
In later sub-sections, we provide details on contemporary work \rev{regarding} personal data stores (PDS), a promising paradigm for enabling individuals to \rev{exercise} autonomy over their personal data, and PETs (such as multi-party computation and differential privacy), technical solutions \rev{for} ensur\rev{ing} data confidentiality. 



\paragraph{Privacy and user behaviours}
\rev{Aside} from \del{privacy concept} discussions \rev{of the concept of privacy}, there is also a related debate about `privacy paradox' \cite{barnes_privacy_2006,barth_privacy_2017,kokolakis_privacy_2017,gerber_explaining_2018}, where users' stated privacy preferences may diverge from their actual behaviour and their willingness to utilise privacy controls, especially in social media and e-commerce contexts. This is also connected to the research on privacy calculus and related topics \cite{gerber_explaining_2018,dinev_extended_2006, acquisti2020secrets, adjerid2018beyond, acquisti_economics_2016, meier_privacy_2024,dienlin_extended_2016}, which considers privacy-related decision-making as a complex mechanism including other factors such as perceived benefits, costs, risks, etc.
In that regard, our work does not directly answer the questions about users' \del{behaviour of} privacy protection \rev{behaviour}. Instead, it focuses on discussing and providing novel technical solutions that can potentially \rev{\textbf{influence}} the benefits, costs\rev{,} and convenience of use, and thus users' perceptions and adoption of user-empowering and/or privacy-enhancing technologies.



\paragraph{Trust}
Similar to privacy, trust is also a multifacet\rev{ed} concept with diverse definitions \cite{corritore_-line_2003,mayer_integrative_1995,thornton_alchemy_2022}, and multiple factors can affect users' perception of trust \cite{kizilcec_how_2016,riegelsberger_mechanics_2005}. Further, perceived trust can also affect user's choice and willingness to use different digital products and services \cite{wei_trust_2022,zheng_user_2018,kim_humans_2023,liu_roles_2022}.
An in-depth discussion on psychological and social mechanisms of trust perception \rev{is} beyond the scope of this work. Our work is based on two commonly agreeable features about trust: 1) trust varies by the subject and the object (including systems), and 2) there are different degrees of trust. More specifically, we allow each individual to express and exercise their own trust preferences, affecting the behaviour of the system. This incorporates additional dimensions not considered by existing literature \rev{on} MPC, where the discussion of threat models as part of security is the main focus (see Section \ref{sec:background_privacy:mpc}), and aligns well with our aim to support user autonomy. We refer to this as \textit{individual-based\rev{,} user-centric trust} (or simply \textit{user-centric trust}), and discuss its implications in system security in Section \ref{sec:solid_mpc:threats}.

\subsection{Personal Data Stores (PDS)}
\label{sec:background:pds}

Personal Data Stores (PDS), such as openPDS \cite{2014openpds}, Solid \cite{2016solid,mansour2016demonstration}, or Databox \cite{2016databox}, champion a decentralised data paradigm by empowering users to retain control over their data within their own data stores, rather than \rev{having it} locked away by large platforms, \rev{thus} fostering user autonomy. \rev{While t}here are subtle technical and design differences between them, \rev{users} generally store their data in a PDS, and applications make requests to the PDS to use (and store) data, \rev{with} user\rev{s} mak\rev{ing the} final decisions about data usage. PDS also offers practical data protection controls, allowing users to establish preferences for data access\rev{,} and \rev{to} audit access to their data. 

While PDS ensures granular control over data access and secure data transmission, privacy faces challenges once an application receives data, especially in scenarios involving collective data use.
For instance, consider a use case which we will demonstrate later in Section \ref{sec:evaluation:solid-mpc:gig-worker}: in a PDS-enabled context, gig workers such as Uber drivers would store their payroll data in their PDS (rather than locked away in their working platform's \rev{data storage}, e.g.~Uber's\del{, data storage}). Now they wish to perform a collective computation task, such as calculating their group's average salary for negotiating better pay rates. A typical \del{PDS} solution would require them to \rev{grant} read permission to their salary data \rev{to a designated individual} (\rev{such as} a worker or \rev{another} delegate)\rev{, who} uses an application to read all \rev{the} data and then compute the average salary. However, this involves raw data transfer\footnote{The network communication channel can be secured by standard means, such as SSL/TLS. However, the issue here is the original ``raw'' data is now copied to the recipient, that application.} from each worker's PDS to \rev{the designated individual} performing computation. Workers have no control once the data transfer is \rev{complete}, and that \rev{individual} can in principle perform arbitrary actions \rev{on} those data. This issue persists \ref{for} all PDS users, all types of data, and all types of collective computation, such as \rev{calculating} census-like data statistics \rev{for} a population, or \rev{deriving} collective insights from health records, as mentioned earlier in Section \ref{sec:intro}.


Some PDS research offers mechanisms to ensure privacy in specific scenarios. For instance, openPDS includes subdivided personal data stores with access controls \rev{and local derivations,} hosted locally or in the cloud \cite{2014openpds}. Similarly, Databox provides a combination of local and remote data stores configured to permit only authorised applications to access personal data \cite{2016databox}. While both emphasise the importance of privacy-preserving aggregation \emph{within individual PDS}, they do not specifically test performing privacy-preserving data aggregation tasks \emph{among a set of users}. Databox \rev{has been} demonstrate\rev{d} to have the potential for distributed privacy-preserving machine learning with \rev{an} implementation of Federated Learning (FL)\del{ among a group of data providers}, orchestrated by a central server \cite{zhao_privacy-preserving_2020}\rev{; \citeauthor{2020meurisch-www} \cite{2020meurisch-www} supported a similar approach with the addition of using secure enclaves like Intel SGX \cite{costan2016intel} for the central server}. However, \rev{such} approach\rev{es} \rev{face the same challenge of }rel\rev{ying} on \rev{one} central server\rev{,} and inherit properties and requirements for FL, \rev{thus being machine-learning-only and} requiring redevelopment when the algorithm or model changes. While openPDS suggests the use of MPC for retrieving aggregated results from multiple PDS \cite{2014openpds}, but to the best of our knowledge, \rev{this has not been implemented or evaluated} \del{did not put the suggestion to test}. 
On the contrary, \cite{anciaux_personal_2019} considered \del{it impractical for} combining PDS with MPC \rev{impractical} \rev{due to the large number of data providers}. 
In general, a mechanism for privacy-preserving data access and processing is needed above and beyond the present features of PDS systems.

\emph{Solid} \cite{2016solid} is a prominent PDS system renowned for its \rev{basis of Web standards and} focus on interoperability, achieved \rev{by} a modular architecture and a clear separation of roles.
Solid allows federation with other users' PDS\rev{s}, known as \emph{Pods}, while preventing vendor lock-in of either PDS hosts or applications. In this work, we explore the implementation of decentralised privacy-preserving computation in a Solid-based architecture rather than Databox or OpenPDS due to its open and standard-based design. We also discuss how \rev{our} approach can be adapted to other PDS systems.

\subsection{Privacy-Preserving Mechanisms}
\label{sec:background_privacy}

Earlier in Section \ref{sec:background:ethical-concept}, we noted that lots of research in computer science focused on the data confidentiality aspect of privacy, which will be the main topic of this subsection. Other than that, there are also some research focusing on information flow management and/or contextual integrity, such as Information Flow Control \cite{pasquier_camflow_2017,myers_decentralized_1997}, various forms of data access/usage control \cite{lazouski_usage_2010,zhao_privacy_2016} (e.g.~Data Terms of Use \cite{zhao_perennial_2024}, ABAC \cite{biswas_label-based_2016,jin_unified_2012}), data governance rules \cite{rui_zhao_draid_2021}, and contextual integrity modelling and application \cite{kokciyan_taking_2022,criado_implicit_2015}. They directly or indirectly relate to \citeauthor{nissenbaum_privacy_2004}'s contextual integrity view of privacy \cite{nissenbaum_privacy_2004,barth_privacy_2006,nissenbaum_contextual_2011}, where privacy is considered to be the appropriate data usage and flow in a given context. We do not \rev{further detail} them because of the difference with the scope of this work\rev{. T}hey express or determine whether information usage should be permitted, \rev{based on} perceived data usage principles\del{ as background}; our work\rev{, on the other hand,} provides a new means for \del{such} data usage principles.

For privacy-preserving mechanisms, there exist several complementary approaches, including data modification, data minimisation, and data encryption~\cite{2021meurisch-survey}. 

\textit{Data modifications} rely on obfuscation- or perturbation-based methods \cite{2014differential,2016deep} to alter or sanitise user data, preventing it from being linked to specific individuals. As a result, it often leads to a difficult trade-off between privacy and utility.

\textit{Data minimisation} strategies aim to achieve optimal computational results by adjusting the computational model and reducing the volume of required data. Federated Learning (FL) is a machine-learning specific approach that trains models over distributed datasets while limiting data movement to central servers, transmitting only model updates \cite{konevcny2016federated,yang2019federated}.
\del{Similarly, }\citeauthor{2020meurisch-www} \del{propose a privacy-preserving platform for personalised ML model training using personal data stored in PDS \cite{2020meurisch-www}, employing secure enclaves like Intel SGX \cite{costan2016intel}}.

\textit{Data encryption} methods utilise encrypted user data to ensure integrity and confidentiality during data sharing.
There are two major data encryption approaches relevant to our work, homomorphic encryption (HE)~\cite{2009fullyhe} and (Secure) Multi-Party Computation (MPC)~\cite{2002smpc}. HE is able to analyse or manipulate encrypted data without revealing the data; however, it is limited by its low computational efficiency and limited operations.
MPC encompasses a class of cryptographic protocols that rely on the secure evaluation of a function over sensitive data shared across multiple parties.
MPC has the benefits of not losing precision and performing any type of computation, providing a promising option for privacy-preserving computation over dispersed data sources. We will briefly present some properties of MPC below, and discuss the challenge of using MPC in a decentralised (PDS) setting.


\paragraph{Secure Multi-Party Computation (MPC)}
\label{sec:background_privacy:mpc}
Given an environment with $n$ parties $P_1\cdots P_n$, their corresponding inputs $x_i\cdots x_n$ and a function $f$, an MPC protocol computes $y = f(x_1\cdots x_n)$ without revealing any input $x_i$ to a party $P_j$ ($i\neq j$).
Traditionally, two security notions have been considered for MPC \cite{oded2009foundations} -- \textit{semi-honest security} and \textit{malicious security}. Protocols with semi-honest security are relatively more efficient and protect against passive attackers that do not deviate from the protocol. In contrast, protocols with malicious security also provide security from attackers that may deviate from the protocol. The number of parties that an attacker could corrupt or compromise is another factor of security in the contexts involving multiple computation parties. Protocols assuming an \textit{honest majority} ensure security under the assumption that fewer than half the parties could be corrupted by an attacker. In a \textit{dishonest majority}, the same assumption does not hold.
MPC protocols are generally realised using a combination of primitives such as oblivious transfer \cite{rabin_how_1981}, garbled circuits (GC) \cite{yao_how_1986} and secret sharing schemes \cite{shamir_how_1979,blakley_safeguarding_1979}.
Here, we focus on additive \cite{bogdanov_sharemind_2008} and Shamir secret sharing \cite{shamir_how_1979}. Understand\rev{ing} the main \rev{discussions and} results in this paper \rev{does not require knowledge of} the details of MPC; therefore, we \rev{omit them for} brevity.

In prior work, MPC has been mostly explored in settings where each computation party has direct access to data \rev{(}i.e.~data providers are computation parties\rev{)}, about the security properties and the performance of protocols and frameworks \cite{keller_mp-spdz_2020,braun_motion_2022}. Work on combining MPC with distributed data sources partially addresses this issue:
\citeauthor{2017secureml} \cite{2017secureml} proposed MPC-based protocols for specific AI algorithms\rev{,} but they only experimented with distributing user data among two non-colluding servers; \citeauthor{2017google} \cite{2017google} proposed an efficient model to securely perform FL over multiple users, but assumed a (single) trustworthy server as with FL in general; \citeauthor{2018deepsecure} \cite{2018deepsecure} use GC to securely perform scalable Deep Learning execution over distributed data from individuals, but is also constrained to the properties (e.g.~performance) of GC; work on Private Set Intersection (PSI) like \cite{freedman_efficient_2004} and \cite{abadi_multi-party_2022} involved many autonomous parties, with or without a central server's help, while intensively exploiting properties of PSI.

Despite their success in increasing the number of data sources, little \rev{attention} has been \rev{paid to the providers of} these computation parties and their relationship to the data subjects, \rev{especially} for \emph{general} MPC support\rev{ing} a wide-range of computation tasks. Essentially, this implies a \emph{centralised trust} of an organisation providing and choosing computing parties. We discuss in Section \ref{sec:decentralised-mpc} how this exhibits problems when applying to a decentralised context such as PDS.

\paragraph{Differential Privacy (DP)}
Output privacy represents a distinct privacy aspect that complements the privacy protection provided by MPC. While MPC ensures that the inputs used in a computation remain undisclosed, output privacy goes a step further by preventing reverse inference based on the revealed computation results. Differential privacy, a formal mathematical concept introduced by~\citeauthor{dwork2006differential}~\cite{dwork2006differential}, plays a crucial role in restricting the disclosure of private information contained in a database when employing a computation algorithm. In simpler terms, an algorithm is considered differentially private if an external observer, upon observing its output, remains unable to determine whether a specific individual's information was utilised during the computation process. In the paper, we show with an example how this complementary notion of privacy can be implemented in our proposed solution.

\section{MPC in Decentralised Settings}
\label{sec:decentralised-mpc}

\subsection{Challenge in Decentralised PDS Context}
\label{sec:mpc-challenges-with-pds}
Existing MPC literature places emphasis on the security-related assumptions about different computation parties, laying a necessary foundation for the Data Privacy (R1) requirement.
However, there is little to no discussion on \emph{who specifies the group of parties and the prescribed security properties}. 

\rev{Usually, e}xisting work focused on a platform-based setting (e.g.~\cite{bell2022distributed,2017google}), so the security property is often assumed as a given because the \textit{platform} will determine the parties and their rights on \del{the} behalf of users.
Thus, \emph{trust} \rev{(and therefore security)} is still centralised \rev{on} the platform, despite using MPC.  This contradicts \rev{the requirements of} User-centric Trust (R3) and User Autonomy (R2)\rev{, leading to a misguided privacy promise}.
\rev{This is the reason we call it the \textbf{`omniscient' view of security}, in which an entity `knows' the security properties of all parties in advance.}

\rev{Nevertheless}, their discussion led to \rev{a} useful observation, which is \rev{the} distinction between
\emph{data providers} and the \emph{App (or App user)}:
a \textbf{data provider} is someone (or someone's PDS) who contributes data to MPC computation;
the \textbf{App} (or \textbf{App user}) is a person or entity who initiates / intends to perform the collective computation over multiple data providers' data.
Naturally, the App user is not necessarily a data provider, and, more importantly, the App user does not necessarily represent the interests or preferences of \emph{all} data providers.
The pitfall of the said \rev{existing} practices result\rev{s from} obfuscating these two roles by naively executing MPC while letting the MPC App developer(s) determine the computation facilities -- by using the App user's machine and/or servers provided by the App developers. However, from the data providers' perspective, these computation parties will not be trustworthy as they can easily collude, unless the data provider fully trusts the App user\rev{. This is rarely the case and negates the need for MPC} \del{which is a rarely true assumption, and also defeats the necessity of MPC} in the first instance.

In an ideal decentralised context with PDS, not only the data storage is decentralised, but also the central trusted party is removed, to ensure data privacy and individual autonomy over data use. Therefore, the platform as conceived in a centralised setting no longer exists, and the MPC App no longer predefines the set of computing facilities and their security features.
Aligning with the ethos of decentralisation and PDS, it is the data owners/providers who should possess the autonomy and ability to administer their trust preferences, and control who has access to their data.

However, because different data providers have different preferences, it is inevitable that they will trust different actors for the relevant actions; also, because they only perceive trust from an individual\del{-based point of view} \rev{perspective}, one data provider's trust (of security properties) is not transferrable to another data provider. Th\rev{is} creates a predicament for MPC because it necessit\rev{at}es a global view of security properties.
Therefore, to enable MPC in such a setting, \rev{an alternative mechanism for establishing permissions is required, along with a mechanism for determining security properties and selecting computing facilities} \del{it requires an alternative mechanism for permission establishment and, importantly, a mechanism for security property determination, in conjunction to computing facility selection}.

Th\rev{is} is the key reason we propose \rev{a shift} from an omniscien\rev{t} view of security properties (\rev{characteristic of} platform-led settings) to an \textbf{individual\rev{-based,} user-centric view of security properties}. We argue that the \rev{assumption} that the system designer knows better than the individuals should be \rev{challenged}, and individual's ability to make decisions for themselves \rev{should be respected}. \rev{C}ollective security properties can \rev{then} be determined and optimised based on individual views, \rev{provided that} appropriate mechanisms are employed, as discussed later with Libertas. That answers three interweaving questions: 

a) Who will carry out the computation; 
b) Why are these parties selected; 
c) What are their security properties?

\subsection{Direct- and Delegated-Decentralised Models}
\label{sec:mpc-models}

\begin{figure}
\centering
\begin{subfigure}{.4\linewidth}
  \centering
  \includegraphics[width=1\linewidth]{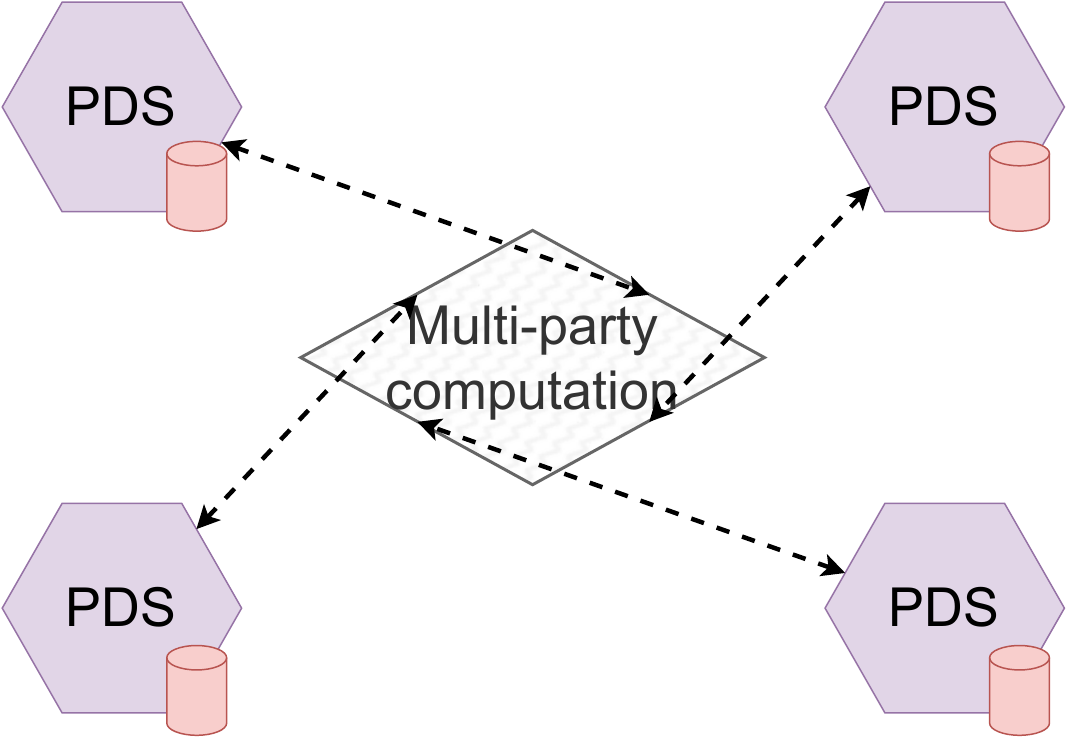}
  \caption{Direct-Decentralised MPC}
  \label{fig:mpc-fully-decentralised}
\end{subfigure}
\hspace{0.1\linewidth}%
\begin{subfigure}{.4\linewidth}
  \centering
  \includegraphics[width=1\linewidth]{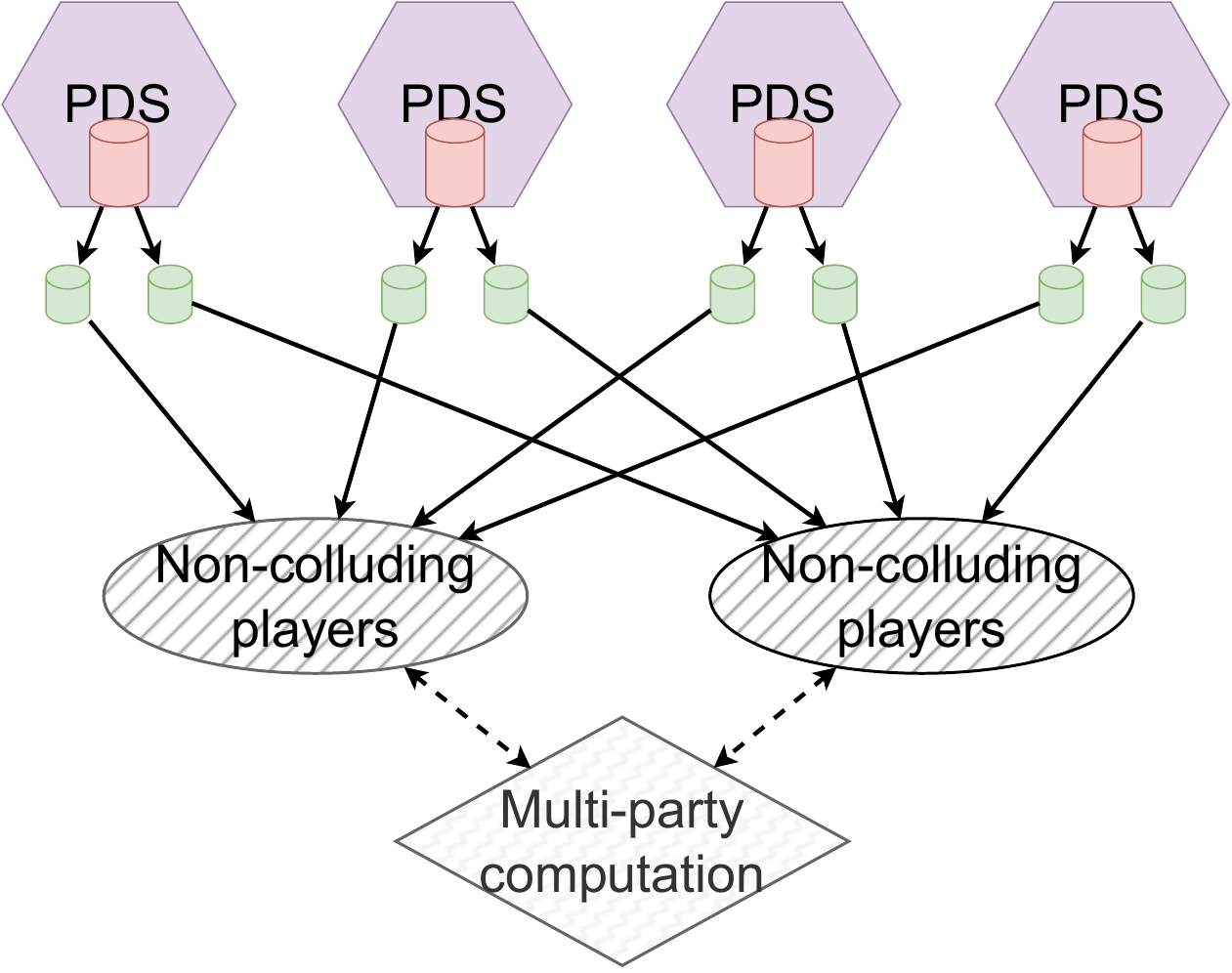}
  \caption{Delegated-Decentralised MPC}
  \label{fig:mpc-semi-decentralised}
\end{subfigure}
\caption{MPC Models in decentralised settings. Data provider is denoted here as PDS.}
\label{fig:mpc-architectures}
\vspace{-.2cm}
\end{figure}

Since our goal is to explore \emph{how} to combine MPC with decentralised PDS, it is useful to revisit the models of MPC. In particular, we focus on \emph{general} MPC protocols that support a wide range of computations, aiming to fulfill the generality (R5) requirement.
Thus, the decentralised computation models for MPC can be classified into two broad categories:
\begin{enumerate}[style=unboxed,leftmargin=0cm]
  \litem{Direct-Decentralised:} (Fig.~\ref{fig:mpc-fully-decentralised}) In this setting, each \emph{data provider} is a computation party (\textbf{player}), following an MPC protocol to carry out secure computation;
  \litem{Delegated-Decentralised:} (Fig.~\ref{fig:mpc-semi-decentralised}) In this setting, \emph{data providers} are different from \emph{players}. Data providers send secret shares of data to these players (such that no single player can make sense of the data independently). The \emph{players} perform the main MPC computation between one another.
\end{enumerate}

The direct-decentralised model reflects the traditional interpretation of MPC -- each computation party (\emph{player}) holds their data and performs computation. The data provider is the same as the player, and the only relationship to discuss is between different players.
On the other hand, the delegated-decentralised model separates the computation parties and the data providers, leading to different relationships.

\subsection{Utilising MPC on PDS}

With this observation, one possible approach is to follow the direct-decentralised model: use the PDS (or a dedicated server for each of them, same below) as the computation parties in MPC. Because each data provider trusts its own PDS, at the minimum a dishonest-majority protocol can be used. Further, if an appropriate punishment or incentive mechanism exists (e.g., retaining social relationships with peers or blacklists), data providers would want to maintain their reputation, and therefore can generally form an honest-majority group.

However, beside the technical requirements (the capability to accept custom computation in PDS), this also involves a performance issue -- the number of data providers (thus \emph{players}) could be very large. As we show in Section \ref{sec:benchmark:mpc-models}, this scales poorly. That is why we would like to avoid this model.

Another approach uses the delegated-decentralised model. The secret sharing of data between the data providers and the computation parties (players) can be realised using a mechanism like the ``\emph{client}'' approach in \cite{damgard_confidential_2017}, which can handle corrupted participants. For \emph{players}, we can expect internal or external services providing agents as player candidates. This provides \rev{the} necessary background design, but \del{we are still yet to answer} the question about trust \rev{remains}: how do we determine the trust relationship\rev{(s)} between the data providers and those computation facilities? Again, we cannot \rev{allow} the App developer to define which facilities to use\del{, and thus there is a need to}\rev{; therefore, we must} incorporate data providers' preferences.

As we will discuss next\rev{,} in Libertas, by allowing users to explicitly express their trust preferences, it is possible to dynamically select computation facilities while respecting users' autonomy. In particular, as we will discuss later in Section \ref{sec:solid_mpc:threats}, a subset of the trusted agents of all data providers can form a non-colluding or honest-majority group (e.g., intersection of trusted agents). This forms the possible basic for faster MPC computation.

The delegated-decentralised model is also expected to scale better than the direct-decentralised model because fewer \emph{players} participate in the computation, which is verified by our benchmark (Section \ref{sec:benchmark:mpc-models}). With this in mind, the next section explains the Libertas architecture we have developed for employing delegated-decentralised MPC with PDS. This approach allows for the utilisation of data providers' preferences for trusted participants, while maintaining compatibility with existing protocols.

\section{Libertas: Architecture for Privacy-Preserving Decentralised Collaborative Computation}
\label{sec:solid_mpc}

\begin{figure*}[h]
    \centering
    \includegraphics[width=0.9\linewidth]{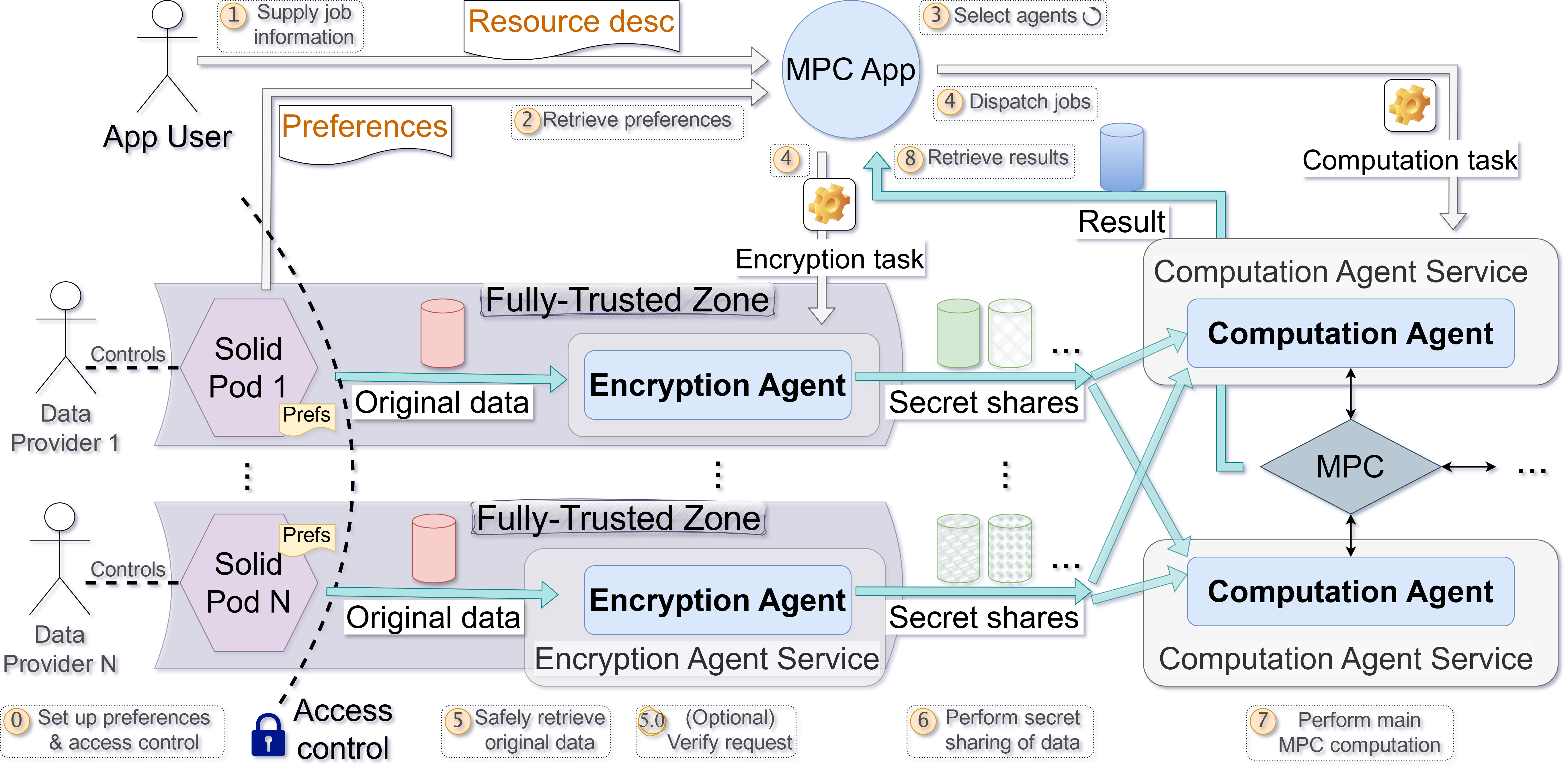}
    \caption{Libertas: A modular architecture for \rev{meaningfully} integrating MPC with Solid. Three dots denote possibly more repetitions. Architectural core components are coloured.}
    \label{fig:solid-mpc-implementation}
\end{figure*}

To address the aforementioned challenges and requirements, we propose an extended architecture based on Solid \cite{2016solid}, called \textbf{Libertas}, illustrated in Figure \ref{fig:solid-mpc-implementation}.
Our architecture \rev{leverages} existing mechanisms such as authorisation and access controls, and is compatible with underlying protocols.
Although we demonstrate a Solid-based architecture, we will also briefly discuss how the proposed architecture can be adapted to other PDSs, thanks to the modularity of the design.

\subsection{Architecture Design}
\paragraph{Core components}
The core components are coloured in Figure \ref{fig:solid-mpc-implementation}.
Users \rev{(data providers)} store their data in their own Solid Pods.
The MPC App initiates the computation and dispatches relevant tasks to the agents who carry out further computation. The \textbf{encryption agents}, acting as the \emph{clients} \cite{damgard_confidential_2017} of MPC, retrieve data from Pods \rev{(Step 5)}, and send secret shares of \del{the} data to the computation agents \rev{(Step 6)}. These \textbf{computation agents}, acting as the \emph{players} of MPC, execute the primary MPC computation amongst one another \rev{(Step 7)}. Finally, the MPC App obtains the results \rev{(Step 8)}. This process differs slightly from the depiction in Figure \ref{fig:mpc-semi-decentralised}, as \del{the} Solid Pods lack computational capability\rev{. Therefore, in addition to the PDS,} \emph{encryption agents} are explicit\rev{ly introduced} \del{in addition to the PDS,} to act as the clients.

\paragraph{Data-provider-centric configuration \rev{(Step 0)}}
Prior to any computation, each data provider creates a \textbf{preference} file, containing their trust preferences regarding the list of trusted actors, especially the encryption and computation agents.\footnote{Further discussion on principles regarding trust and agent selection is provided in the subsequent section. Additional preference files associated with data resources are used for additional security guarantee, also discussed later.}
More precisely, it contains the \textit{services} providing such agents, which will be elucidated later in the discussion on \textit{separation of duty}. As files are protected by Solid's access control mechanism (Web Access Control \cite{berners-lee_web_2022}), the data provider \rev{must} also \del{needs to} grant relevant permissions. By doing so, the provider makes the preference file readable to the MPC App or the App user\footnote{Because at each time, the App and App user authenticate themselves together. For simplicity, we will only refer to them as the MPC App.} and grants the trusted encryption agents (identified by their WebID\rev{s} \cite{virginia_balseiro_solid_2022}) read \rev{access} to their data. These steps only need to be performed once and can remain the same for all future MPC tasks if the preference does not change. If the data provider chooses to revoke trust from anyone, they can simply withdraw access permission\rev{s} from their Solid Pod and/or removing them from the preference file\rev{, as appropriate}.

\paragraph{App usage \rev{(Step 1 \& 4)}}
The App user submits a \textbf{resource description}, containing a list of data (resources) and the relevant \emph{preferences} from each data provider. Assuming the appropriate permissions are granted, the MPC App reads the preference files, identifies the encryption \del{agents} and computation agents (to be discussed later), dispatches relevant directives (and MPC tasks) to them, awaits the completion of the computation, and retrieves the result.

The \textbf{encryption task} describes how to \rev{secretly} share data with each player, while the \textbf{computation task} specifies how to execute the main computation among the computation agents. Both the encryption task and the computation task should be accompanied by the list of computation agents (instances of such agents obtained from the services) to facilitate connection establishment. We utilise MP-SPDZ \cite{keller_mp-spdz_2020} as the MPC framework, leveraging its supported \emph{client} mechanism, which is based on a variant of the SPDZ protocol~\cite{damgard_confidential_2017}, for the secret sharing of data. In our implementation, the tasks are transmitted as source code, and the agents will (compile and) execute them upon receipt. Additionally, the App can customise various aspects such as the number of computation agents (and methods to determine them), MPC protocol, parameters, etc., as MP-SPDZ supports a wide range of protocols and parameters.

\paragraph{Trust and agent selection \rev{(Step 2 \& 3)}}

The preference files provided by data providers serve as the primary source for the MPC App to select relevant agents (and for encryption agents for further verification, for additional security). In our prototype, each preference file includes a list of trusted encryption agents and a separate list of trusted computation agents.

It is important to note that the trust requirements for encryption agents and computation agents differ: encryption agents must be fully trustworthy \rev{from} the data provider's perspective, while computation agents only need to be semi-honest and non-colluding. This distinction arises because encryption agents have access to raw data, whereas computation agents receive \rev{only secret} shares of data. Ensuring the non-collusion of computation agents is crucial\rev{:} if all of them collude (or a significant portion of them collude, depending on the protocol), they can combine their shares and reconstruct the original data. These agent services can be provided by public or private entities. For instance, the PDS host may deploy an encryption agent service alongside the PDS service itself, making it relatively straightforward to establish trust with the data providers.

Choosing the encryption agent for each data provider is straightforward for the MPC App: simply select one of the trusted agents specified by that data provider. It is permissible to choose the same encryption agent (service) for different data providers. However, selecting computation agents requires more careful consideration. Various approaches can be adopted for this purpose, and we have implemented two extreme versions: taking (a subset of) the intersection of the trusted computation agents of all data providers, and taking the union. Further details are discussed in Section \ref{sec:solid_mpc:threats}.

\subsection{Additional Properties}
\label{sec:solid_mpc:additional_property}
\paragraph{Separation of duty}
The Libertas architecture effectively partitions the responsibilities of various stakeholders, resulting in reusable components across multiple tasks. Both types of agents, encryption and computation, can be automatically repurposed by different MPC applications because they solely provide computing (and data fetching) \emph{facility} rather than \emph{logic}. This suggests the existence of public or private, free or paid, agent services from which data providers can make selections. The actual computation is instantiated by the MPC App, which furnishes the logic (and parameters) for the computation. \rev{Because} the agent services operate independently of the App, each App can seamlessly utilise existing agents (agent services) specified in the \rev{data providers'} preferences.

Furthermore, as the App user may not necessarily be a data provider, this setup affords data providers the flexibility to authorise third parties to utilise the data for legitimate computations, such as platforms, worker unions, or government entities. Access controls are implemented to prevent unauthorised access, and the security mechanisms and secure protocols of MPC safeguard the raw data from exposure to the App user.

\paragraph{App verification \rev{(Step 5.0)}}
The encryption agent may additionally perform \emph{App verification} to enhance the security properties of the system. To support this, the data provider must express the list of trusted Apps in a resource (called \emph{trusted actors}), making it readable to trusted encryption agents, and refer to the URI of \emph{trusted actors} in the \emph{description resource} of the data \cite{2021solidprotocol}. Upon receiving computation request, the encryption agent retrieves the \emph{description resource} and then the \emph{trusted actors}; it also attests the identity of the App by checking the request headers (following Solid-OIDC \cite{coburn_solidoidc_2022}); then, it verifies if the App is within the \emph{trusted actors}. This process both verifies the real identity of the App, and also verifies the App is within the trusted actors.

\paragraph{Variants}

As outlined earlier, the Libertas architecture seamlessly integrates with the existing Solid protocol and leverages its established mechanisms. From a broader perspective, this architecture is not confined to a specific ecosystem or framework for either the PDS or MPC framework, but rather relies on functional properties such as the presence of an access control mechanism. This flexibility allows for its adaptation across various systems.

To accommodate different (PDS) systems, the different modules of Libertas may be merged with them, yielding derived architectures with distinct trade-offs. For instance, merging the encryption agent with the data store (PDS) simplifies the user's task of identifying and assigning trust to encryption agents. However, this approach may complicate the PDS service, potentially deviating from its original protocols. This exploration could be particularly relevant for PDS systems with computation capabilities, like openPDS. Alternatively, integrating computation agents with the PDS services and randomly sampling a subset of them via the MPC App could be another avenue to explore.

Moreover, the flexibility extends to the MPC framework, allowing for seamless substitution while preserving the architecture's integrity.

\subsection{Threat Model}
\label{sec:solid_mpc:threats}

In Libertas, four major parties play essential roles: the Data Provider (DP), the App (including App user), the Encryption Agent (EA), and the Computation Agent (CA). This presents a distinct threat dynamics compared to existing MPC literature, where only EA and/or CA are considered (Section \ref{sec:background_privacy:mpc}). This section discusses that in depth.

\subsubsection{Assumptions}

As a basis for discussion, here are the general assumptions:

\begin{enumerate}
    \item All network transmission is secure, for example through SSL/TLS.
    \item Output data from Libertas are not a source of privacy concerns (e.g. through Differential Privacy as discussed in evaluation).
    \item User's PDS/Pod is fully trustworthy.
    \item The implementation and theory of the authorisation and authentication mechanisms come without flaws.
    \item Hosting machine and software stack do not involve additional security concerns.
    \item Each party in the framework has a unique identifier to be identified.
\end{enumerate}
Besides that, since the motivation for Libertas is to support user autonomy (while also achieving scalable collective computation), we also assume the DPs are semi-honest and non-colluding, and they make sensible decisions (e.g.~to their preference documents and access control settings) to the best of their knowledge. That means, DPs may choose trusted agents for their own benefits but not others.

It is worth emphasising that Libertas utilises individual-based\rev{,} user-centric trust: each DP only makes decisions for himself/herself, and they do not need to know or consider the decisions of other DPs. This individual-based view is a key factor for the discussion below, as it is the only source of information in a user-autonomous setting. We will not repeat this at all times in the discussion below.

\subsubsection{Threat from Encryption Agent}

An EA \rev{can} access all data that it \rev{has been granted read} permission to\rev{; therefore,} we require every EA to be fully trustworthy. This is why it may be impeding to combine the EA with the PDS service, as discussed above in \emph{variants} (see also discussion below about the potential threat from the App).
If an EA is compromised, all data it has access to are at the risk of privacy leak.
The only countermeasure relies on avoiding the EAs from being able to enumerate the resources in users' Pod{s}, such as by denying access from trusted EAs from a parent/ancestor container (but allowing access to the data resource file). However, that only protects data that have not been accessed by the EA.

As we have employed access control to the data resources, untrusted EAs will not be able to gain access to the data.

\subsubsection{Threat from App}

As an App will dispatch tasks to EAs and CAs, and is responsible for the selection of EAs and CAs, we require it to be semi-honest. That means the App will correctly follow the protocol, including retrieving preference files, selecting EAs and CAs, and dispatching tasks.
Since the App has no access to raw data, and we have assumed the computation result is not sensitive, there will be no threat concern for a semi-honest App to obtain sensitive data, such as the raw data of each DP.

An untrusted App may send random requests (dispatching tasks) to EAs, in the hope that it may opportunistically choose the right combination of data resource and EAs. In this case, it will set up and instruct EAs to perform secret sharing to contaminated CAs controlled by itself (see also the discussion of threats from CAs below).
In this case, the \emph{App verification} procedure discussed in Section \ref{sec:solid_mpc:additional_property} can eliminate such threats, as the EA receiving the request will be able to identify that the App is not trusted by the data provider for this specific resource, and ignore the request.

For this reason, the vanilla variant of Libertas prefers EAs to be provided separately from PDS services, as this reduces the possibility of an untrusted App choosing a sensible combination of data resources (, data providers) and EAs on the Internet, by a factor of the possible EA services on the Internet. \rev{However,} we also proposed the \emph{App verification} mechanism for variants \rev{that} combine EAs with PDS to simplify the trust selection (given that both PDS and EA \rev{share the same} trustworthy \rev{requirement})\rev{, or implementers who prefer a more cautious approach}.

\paragraph{Threat from compromised App}
If a trusted App becomes malicious, it may perform one of the following to form a threat to data privacy:
\begin{enumerate}
    \item Send malicious computation tasks to the CAs (e.g.~revealing input directly as output);
    \item Instruct EAs to use malicious CAs for computation, and gain access to input from these CAs (this includes selecting MPC protocols with inappropriate security guarantees).
\end{enumerate}

For case 1, the DP can assign trustworthiness to computation tasks, such as by digital signatures or verifiable credentials, and share that with the App in advance. The CA will need to verify them \rev{upon} receiving the task. For example, when sending the computation job (to the CAs), the App \rev{must also} send the digital signatures of all DPs (confirming approval) for this job (code); the EA will send the public key of the DP to the CAs; each CA verifies all the signatures are valid, and then proceed with computation. Note the assumption that (at least one) CAs are trustworthy, as to be discussed in the next section.

For case 2, a countermeasure is to verify the appropriateness of the chosen MPC protocol and the chosen CAs. As \del{to be} discussed in the next section, not all chosen CAs need \del{to} be trusted by the DP, but there is an appropriateness between the (trustworthiness of) chosen CAs and the MPC protocol. Since CAs will be established communication from EAs and thus named in the job dispatch from the App, the EA (of each individual DP) can verify the trustworthiness of the chosen CAs. For this, the EA will need to 1) verify the amount of chosen CAs that are trusted by the DP, 2) calculate the allowed MPC protocols\footnote{The security \rev{property} of each MPC protocol is well-known public information, so the calculation here is to simply select the set of MPC protocols that fits with the security property fact -- the ratio between trusted CAs and chosen CAs. To strengthen security or further restrict the range, the DP may specify a list of allowed MPC protocols, and the EA calculates from this set.}, and 3) send that to the CAs; each CA will 4) verify if the chosen MPC protocol is within the list, and abort if otherwise. This works \rev{on} the assumption that at least one chosen CA is \rev{among the DP's} trusted CAs; if otherwise and that is a concern (see Section \ref{sec:threat:ca_from_union} for a scenario where this may not be a concern), the EA can abort the computation \rev{immediately upon discovering} this (at step 1).

\subsubsection{Threat from Computation Agent}

We require the CAs to be semi-honest and non-colluding \emph{to the DP nominating them in the preference file}. However, \rev{unlike} EAs, CAs are shared across all DPs \rev{for each computation run, making} their security propert\rev{ies} more complex.

The computation run will need to choose an MPC protocol with appropriate security guarantees given the security properties of the chosen agents to avoid data privacy risks. \rev{While} stronger protocols \rev{offer} theoretically higher privacy guarantees\del{, for the sake of performance}, that is unnecessary \rev{for performance reasons. T}he chosen protocol should have the same security assumption as the security properties of the chosen agents.
Because Libertas uses the \emph{client} mechanism \cite{damgard_confidential_2017} for secret-share of data from EAs to CAs, privacy is protected even if all clients (EAs) are corrupted and all-but-one players (CAs) are corrupted. In the meantime, if all CAs are corrupted, no privacy is guaranteed between CAs (not considering EAs) regardless of the chosen MPC protocol. Thus, the secret-share mechanism will not be the privacy bottleneck, and we will not discuss it below.

For agent selection, in the best case, if the agent selection algorithm is \emph{subset}, which takes the intersection of trusted CAs from the preference files of all DPs, the chosen CAs will be semi-honest and non-colluding to all DPs. In this case, employing a semi-honest MPC protocol will suffice, such as Shamir.

Alternative mechanisms \rev{are} needed \rev{when an insufficient number} (but still some) \rev{of suitable} CAs \rev{are} found in the subset, because non-common CAs are not necessarily semi-honest or non-colluding \rev{with respect to} the \rev{other} DPs. In these cases, to accommodate the required number of CAs, the selection algorithm will need to choose enough number of CAs from the non-intersecting union of CAs, leading to non-trusted CAs in the final list of chosen CAs.
Therefore, with the increase of non-trusted CAs, MPC protocols with stronger security guarantees may be needed, up until a covert (e.g.~ChaiGear or CowGear \cite{keller_overdrive_2018}) or malicious (e.g.~MASCOT \cite{keller_mascot_2016}) protocol, to detect (and abort) and/or tolerate dishonest behaviours in this context, at the cost of performance. For example, if only one common CA exists, we can use protocols like MASCOT for dishonest-majority security.

\subsubsection{Special Case for Lack of Common\rev{ly} Trusted Computation Agents}
\label{sec:threat:ca_from_union}

In the worst case, no common trusted CAs exist, and thus each individual DP cannot get assured a security property of the chosen CAs. In that case, either the computation is not possible, or the DPs have to accept a risk. We note a special case where an acceptable slight risk is exhibited to perform the computation in this scenario: if we \emph{randomly} \rev{select} from the union of all trusted CAs of every DP, and we use a malicious-majority protocol, such as MASCOT, privacy can still be guaranteed \rev{provided} that not all the computation agents are corrupted.

Assuming each DP provides the same number of distinct CAs, this risk probability is $P_{risk} = \prod_{i=0}^{m-1}\frac{k-i}{n-i} < (\frac{k}{n})^m$ for random choice, where $n$ is the number of agents in the union of trusted CAs of all DPs, $k$ is the number of corrupted agents in the union and $m$ is the number of chosen CAs.
With a sufficiently large group of DPs (thus large $n$) and a limited $k$, this probability is close to zero. This implies that as long as the majority of DPs are honest, the risk is very low and is acceptable if \rev{possibility of} computation is \rev{more} important.


Of course, future work can explore alternative selection algorithms for better suitability in different scenarios, especially utilising context-specific information, for an efficient balance between security and performance.

\section{Empirical Evaluation}
\label{sec:evaluation}

In this section, we present a series of benchmarks on various tasks. Since our work fills a gap in the application of MPC rather than proposing new MPC protocols, and we utilise an off-the-shelf MPC framework, comparing performance against other works as a baseline may not be meaningful. Instead, we concentrate on scalability patterns concerning a distinctive feature in the decentralised PDS context: the number of data providers.

\subsection{Scalability of the Two MPC Models}
\label{sec:benchmark:mpc-models}

Using the same framework, MP-SDPZ~\cite{keller_mp-spdz_2020}, we conducted a scalability comparison between two MPC models: direct-decentralised MPC and delegated-decentralised MPC.

In the benchmarks, each data provider possesses an array of data and employs MPC to compute an element-wise operation across all data, culminating in the summation of these results.\footnote{For instance, in a two-player setting with multiplication as the operation, this is akin to computing the dot-product.} Our benchmarks encompass various computation operations (such as sum and multiplication), parameters (such as array sizes and numbers of parties), and protocols (such as Shamir and MASCOT) in the two models, utilising a server with 2x 8-core Intel E5-2650v2 (2.6GHz) CPUs and 48GB RAM. The primary factors/metrics recorded include time, rounds of communications, and data transmission.

\begin{figure}
    \centering
    \begin{subfigure}[sub]{.3\linewidth}
      \centering
      \includegraphics[width=\linewidth]{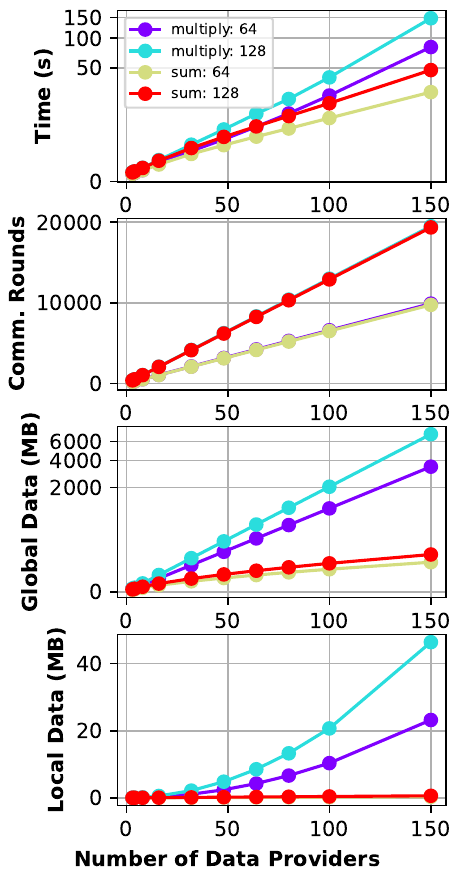}
      \caption{Direct-decentralised MPC. Time and global data are in \textbf{cubic-root scale}. Legend shows the computation \& array size. \newline}
    \end{subfigure}
    \hspace{0.025\linewidth}
    \begin{subfigure}[sub]{.3\linewidth}
      \centering
      \includegraphics[width=\linewidth]{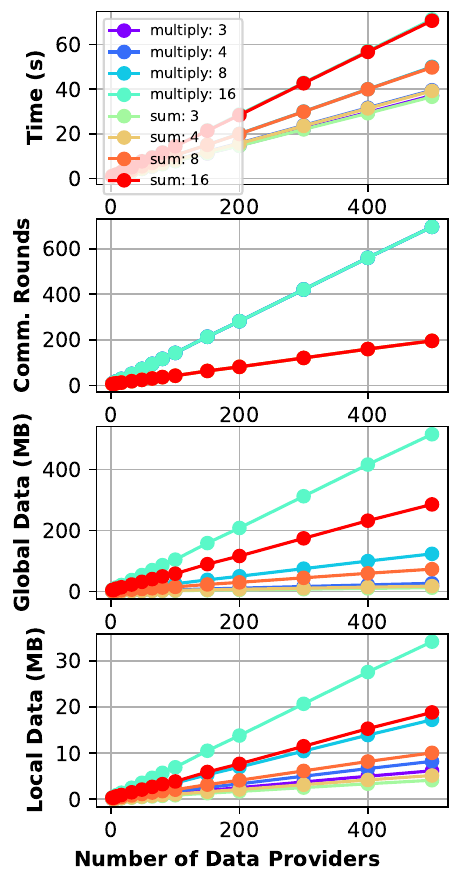}
      \caption{Delegated-decentralised MPC. Legend shows the computation and number of players. Array size is fixed to 128. \newline}
    \end{subfigure}
    \hspace{0.025\linewidth}
    \begin{subfigure}[sub]{.3\linewidth}
      \centering
      \includegraphics[width=\linewidth]{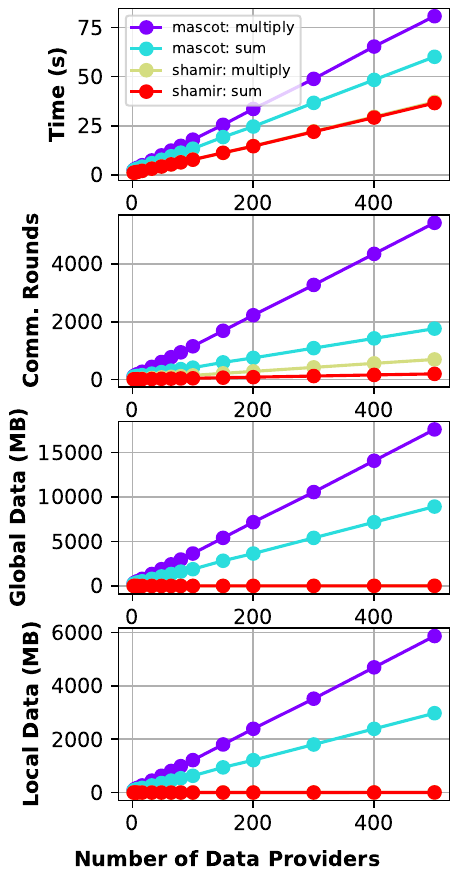}
      \caption{Protocol comparison in delegated-decentralised MPC, with 3 players and array size 128. Legend shows protocol \& computation.}
    \end{subfigure}
    \caption{Benchmark of two MPC models in different settings.}
    \label{fig:mpc_benchmark}
\end{figure}

The results are depicted in Figure \ref{fig:mpc_benchmark}. In summary, we observed and confirmed the following:
\begin{itemize}
   \item In all settings, computation cost of delegated-decentralised MPC grows linearly with the number of data providers, while that of direct-decentralised MPC grows polynomially.\footnote{Quadratically for sum; cubically for multiplication.}
   \item Cost for more complex computation (multiplication) grows faster than simpler computation (sum).
    Sometimes this is in magnitudes of difference (i.e.~multiplication vs sum for direct-decentralised).
   \item Protocols with stronger security guarantees require significantly more resources, and are more costly than more complex computational operations.
    For example, sum operation in MASCOT is more costly than multiplication in Shamir.
\end{itemize}

\noindent From the results, we conclude that the delegated-decentralised model is more appropriate for a significant decentralised context such as Solid. 
This can be intuitively explained: increasing the number of data providers equals to increasing players in a direct-decentralised model, thus leading to more rounds of transmissions, and a higher number of communications and data transmission in each round; while more data providers in a delegated-decentralised model results in more clients, but does not change the number of players inter-communicating to perform the main computation.

We also observe that each server can handle a sufficient number of players/clients, without impacting performance. Although we did not specifically test the maximum capacity of the server hardware, we stopped at the shown numbers (150 players for direct-decentralised and 500 clients for delegated-decentralised) because of soft operating system limits (esp.~number of file descriptors). Th\rev{e shown results} demonstrate the feasibility of running dedicated servers, for the \textit{agents} proposed in the Libertas architecture.

Furthermore, in our benchmark, the time required for even basic operations cannot be ignored, especially for protocols with stronger security guarantees. This may pose challenges for tasks requiring real-time processing in large decentralised contexts. However, it is important to note that we did not fine-tune parameters, and the time includes initialisation and connection establishment between all parties, which can impact \rev{the reported} performance. Further research in MPC may lead to performance improvements in the future, which can be directly integrated into Libertas, such as a new protocol. Conversely, for tasks that do not require real-time processing, this is less of a concern, particularly for tasks where privacy outweighs efficiency, such as medical records or highly sensitive financial data. The upcoming use cases discussed below provide additional insights in this regard.

\subsection{Libertas Evaluation in Real-World Use-Cases}
\label{sec:evaluation:solid-mpc}
To demonstrate how Libertas can deliver impact, we assess its ability to support realistic use cases, such as gig worker empowerment, and differentially private synthetic dataset generation from decentralised data sources.

\subsubsection{Gig Worker Empowerment}
\label{sec:evaluation:solid-mpc:gig-worker}

Gig workers are vulnerable to unfair treatment by their work platforms \cite{calacci_organizing_2022,sannon_more_2022}. They could benefit from gaining control over their own data and conducting statistical analysis on data aggregation, a practice repeatedly mentioned in literature \cite{calacci_access_2023,yao_together_2021,calacci_bargaining_2022}. They also informally perform this through unsecured social media channels and private text groups, demanding substantial effort from moderators and entail placing significant trust in fellow group members or platforms, also risking the exposure of sensitive information \cite{yao_together_2021}.
Several recent studies have highlighted that the preservation of workers' personal privacy is critical for their uptake of any new paradigm, as sharing salary information or work patterns may jeopardise their earnings or competitiveness amongst other workers \cite{zhang_algorithmic_2022,open_data_institute_perceptions_2022}.

The first evaluation we present discusses how Libertas can facilitate gig workers by providing a more flexible, agile and privacy-preserving framework to empower the gig workers, without involving a single source of trust, reducing risks in existing approaches discussed above. We first present how Libertas can support the practice, including how the requirements of Libertas can be achieved by real-world power dynamics, and then present the technical solution aspect, especially the experimental conditions.

\paragraph{New Norm with Libertas}
By adopting Libertas, gig workers will store each individual's data in their own Pods, and execute the computation by themselves or delegates with \emph{computation rights} (but not \emph{data access rights}) to the data.
The fact that they may already trust a data intermediary can accelerate the adoption, because the original actors such as data intermediaries can still provide multiple functions, such as providing Pod storage, data analysis, and agents for computation.
This may seem unnecessary if believing everyone to trust the same party, but immediately becomes useful if otherwise: because of the clear and agile design of Libertas, any single gig worker can choose different providers for each part, without necessarily impeding the collective computation, which is extremely difficult in existing practices.

For example, a gig worker may choose to\rev{:} 1) store the data in a Pod at the work platform\rev{;} 2) use a self-hosted EA\rev{;} 3) permit a worker union for the analysis App\rev{;} and/or 4) use CAs provided by different institutions, \rev{according to} their individual preferences. In particular, real-world power dynamics can lead to sensible choices of such CAs, where different institutions have an intrinsic tension \rev{that prevents them from} cooperat\rev{ing} with one another. For example, workers \rev{might} choose one CA from the public sector (e.g.~\textit{transportation authority}), one CA from their platform (e.g.~\textit{Uber}), and one CA from \rev{a} worker union or data intermediary (e.g.~\textit{App Drivers \& Couriers Union}). Their well-known \rev{reputation and structural tension} result in a high \rev{likelihood} for the majority of workers \rev{to} trust them as CAs; even if not all workers trust the same CAs, as discussed previously, the computation can still be performed without privacy concerns.

\paragraph{Experimental Design}
We present an experiment to assess the technical feasibility and scalability pattern of Libertas in such a context. As a use case, it tackles the wage discrimination often faced by gig workers \cite{dubal_algorithmic_2023}, by providing insights of average wage across all gig workers from individually owned private income data.
In the experiments, we assume that each worker is equipped with a Pod, storing their hourly income (simulated by assigning random values), and they have chosen appropriate actors for the collective computation using Libertas. We implemented an "average wage" \rev{MPC} circuit \rev{in MP-SPDZ}, calculating the average wage from individual earnings stored in their Pods. Experiments were conducted with varying numbers of data providers (10 - 1000) to discern the scalability pattern.

\subsubsection{Synthetic Dataset Generation with Differential Privacy}
\label{sec:evaluation:solid-mpc:dp}
As discussed in Section~\ref{sec:background_privacy}, while MPC ensures that the inputs used in a computation remain undisclosed, no privacy protection is provided for the computation results. Orthogonally, Differential Privacy \cite{dwork_calibrating_2006,dwork_differential_2008} \rev{provides} a series of mechanisms to protect the output from being used to identify information about input. It complements the input privacy of MPC, and thus enables an interesting use case for Libertas for synthetic data generation. In this use case, we delve further by discussing the benefits and technical aspects of performing differential privacy on Libertas.

Formally, differential privacy provides a formal specification for a randomised function $\mathcal{M}$ that takes input $D$ (from an input space $\mathcal{D}$) and produces output $O$ \rev{(}from space $\mathcal{O}$\rev{,} not necessar\rev{ily} different from $\mathcal{D}$) -- the function $\mathcal{M}$ is said to provide $\epsilon$-Differential Privacy if for all \emph{neighbouring} datasets\footnote{Neighbouring dataset means $D$ and $D'$ differ in at most one entry, e.g.~$D$ has one more entry than $D'$.} (of $D$) $D' \in \mathcal{D}$ and for all subsets $\mathcal{S} \subseteq \mathcal{O}$, such that
\[
Pr[\mathcal{M}(D) \in \mathcal{S}] \leq e^\epsilon \cdot Pr[\mathcal{M}(D') \in \mathcal{S}]
\]
To put it differently, the function, or algorithm, $\mathcal{M}$ produces indistinguishable outputs for two datasets that only differ by a single entry, bounded by the privacy parameter $\epsilon$.
Consequently, it safeguards against using the output to deduce whether a specific data record exists in the dataset or if a particular data provider contributed to the computation (assuming each provider contributes only one data record).

Differential privacy mechanisms can be used to generate synthetic data that mimic the statistical properties of real-world data while minimising privacy risks, offering a means to balance the need for data-driven insights and open availability of data with the imperative of protecting individual privacy \cite{jordonSyntheticDataWhat2022}. Beyond privacy, synthetic data approaches are also being actively explored to overcome the limitations and shortcomings of real data for building more robust and fair artificial intelligence \cite{emamPracticalSyntheticData2020a,liFairGANGANsbasedFairnessaware2022}. Note that this is not to imply that differentially private synthetic data is a silver bullet for data privacy (let aside the differences in scope of privacy, as discussed in Section \ref{sec:background:ethical-concept}). \rev{Indeed,} there are several known limitations of synthetic data due to the fundamental trade-off between accuracy and privacy risks \cite{stadler2022synthetic}. But depending on the use case and context, such techniques \rev{can} be very useful~\cite{calcraftaccelerating,emam_practical_2020, jordon_synthetic_2022,bowen2019comparative} in practice.

However, having access to raw data to generate high quality synthetic data is often a challenge, partially because of the common assumption of the existence of a central curator, in most synthetic data generation approaches. This is similar to the gig worker empowerment scenario\rev{, where the impractical} assumption \rev{of} requir\rev{ing} a diverse set of contributors to trust a \rev{common} central curator may lead to reduced or \rev{even} dishonest contributions, \rev{consequently} lowering the quality of data \cite{mcsherryMechanismDesignDifferential2007a}.

Using Libertas to coordinate the data providers provides a novel paradigm of synthetic data generation. To use Libertas, users (data providers) need \rev{only} express the relevant trust information and assign access control in their Pods, all \rev{on} an individual \rev{basis;} and the \rev{architecture} automatically guarantee\rev{s the rest}. It offers several benefits:

\begin{itemize}
    \item Data providers do not need to coordinate with each other in advance or simultaneously with a single data curator, lowering the precondition requirements of trust obtaining and assignment.
    \item By ensuring privacy protection at both the input and output ends, users \rev{would} feel more confident participating in synthetic data generation. This can lead to the creation of higher-quality and privacy-friendly open synthetic data sets for the common good.
    \item The use of costly MPC is minimised, as it is employed only \textbf{once} during the generation of synthetic data. Subsequently, the synthetic data can be used for running queries and analysis in a privacy-friendly manner \textbf{without} the \rev{continuous} need for MPC \del{operations directly on sensitive personal data stored in personal data stores}. This approach is significantly more scalable than employing costly MPC for every query and analysis \rev{on sensitive personal data stored in personal data stores}. Additionally, different synthetic data generation algorithms provide certain guarantees on the accuracy of queries, enhancing usability of the data.
\end{itemize}

\paragraph{Experimental Design}
We implemented the classic Multiplicative Weights and Exponential Mechanism (MWEM) algorithm \cite{hardt_simple_2012} (see Appendix \ref{sec:appendix:mwem} for more details of MWEM algorithm) for differential privacy as an MPC circuit,\footnote{We implemented MWEM of 1-D dataset (integers), taking the final distribution as output, and do not perform mini-iterations during the multiplicative weights update step. With the produced distribution, one can sample a synthetic dataset.} and initiated computation through our prototype App.
In the experiment\rev{s}, randomly sampled data and preferences are distributed across various resources under different containers, simulating different Pods within the decentralised architecture.

We conducted experiments under the following settings with varied number of data providers from 10 to 1000:
\begin{enumerate*}
    \item[(\textbf{\rev{Setting} 1})] Fixed total number of data points (10,000), evenly distributed among data providers;
    \item[(\textbf{\rev{Setting} 2})] Fixed number of data points per provider (100).
\end{enumerate*}
The MPC MWEM circuit transformed the data into a histogram using a fixed number of bins (10). The MWEM algorithm was executed with 60 randomly pre-generated queries for 30 iterations ($T=30$) and an epsilon value of 1 ($\epsilon = 1$). Additionally, we conducted another set of benchmarks with a simple yet proven efficient optimisation (\textbf{Setting 3}): allowing clients to create local histograms and send these histograms instead of having players create histograms.

\subsubsection{Experimental Settings}

For both scenarios, we maintained the following common settings:
\begin{itemize}
    \item Deployment of 3 computation agent servers and 1 encryption agent server\rev{,} interconnected over a (virtual) LAN.
    \item Each computation agent resides on its own server, aligning with the non-colluding requirement discussed earlier.
    \item The agents utilise self-signed SSL certificates to ensure secure data transmission between them.
    \item All servers are equipped with 2x Quad-core Intel E5520 (2.2GHz) CPUs and 12GB of RAM.
    \item Data is hosted on a Solid server running Community Solid Server implementation \cite{inrupt_community_2023}, situated in the same data centre as the agent servers but not directly linked via (virtual) LAN.
    \item We employed the Shamir protocol, utilising the default parameters from MP-SPDZ.

\end{itemize}

We recorded the relevant factors for computation on the 1st player. Specifically, we noted the time for the entire job (\emph{full-time}),\footnote{To be precise, this duration spans from when the players/CAs connected to each other until the completion of the computation.} as well as the time after all connections were established (\emph{comp-time}).\footnote{This excludes the time required for encryption agents to prepare and download data, \rev{and to} establish connections with computation agents.} Each task was repeated at least 10 times to obtain \rev{an} average result, \rev{thereby} mitigating the impact of random fluctuations.

\subsubsection{Results}

\begin{figure*}[h]
    \centering
    \includegraphics[width=\linewidth]{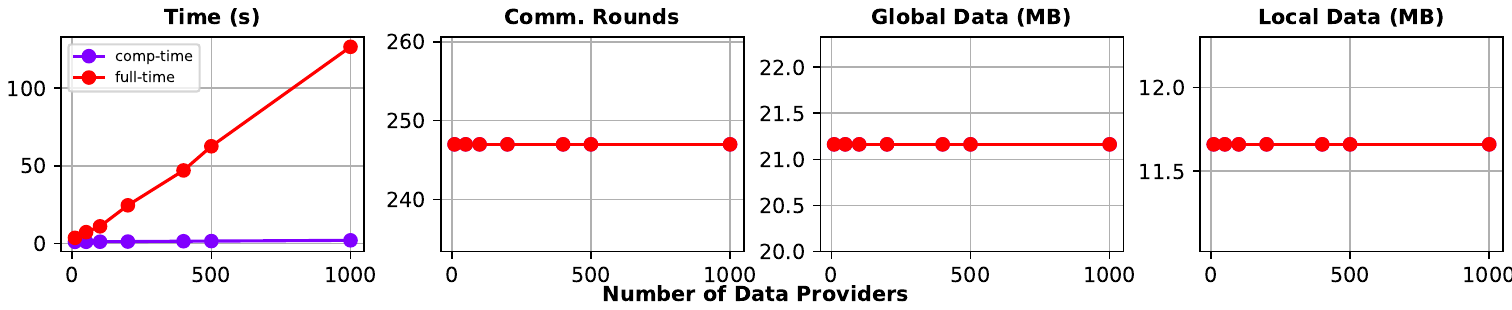}
    \caption{Results for average wage computation in Libertas in the gig workers scenario.}
    \label{fig:solidmpc_average_wage}
\end{figure*}
\begin{figure*}[h]
    \centering
    \includegraphics[width=\linewidth]{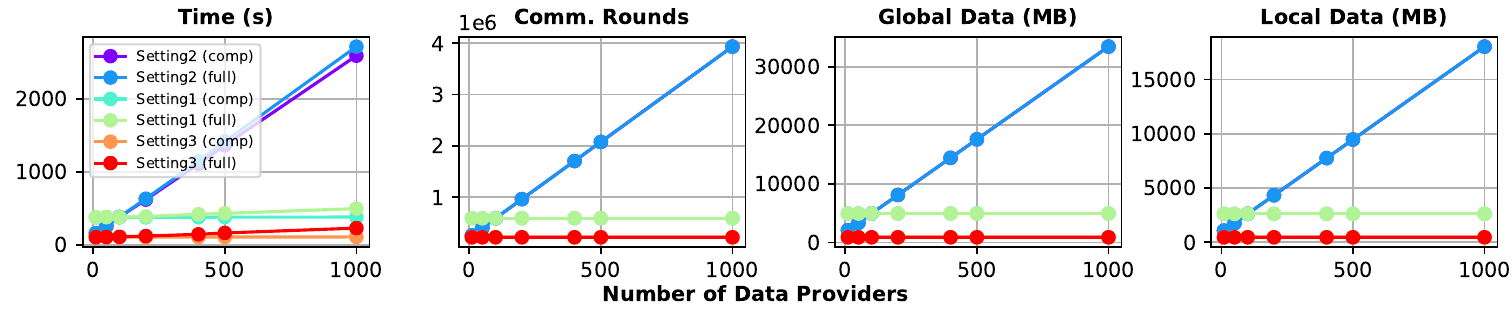}
    \caption{Results for differentially-private synthetic data generation (MWEM) computation in Libertas.}
    \label{fig:solidmpc_mwem_combined}
\end{figure*}

Figures \ref{fig:solidmpc_average_wage} and \ref{fig:solidmpc_mwem_combined} display the experimental results across various settings. It is evident that all factors exhibit a linear growth trend with the increase in the number of data providers. This aligns with the linear trend observed in the platform-agnostic benchmark for delegated-decentralised MPC (Section \ref{sec:benchmark:mpc-models}), affirming the overall scalability of the architecture.

Zooming into Figure \ref{fig:solidmpc_mwem_combined}, a comparison between Setting 1 (green lines) and Setting 2 (blue lines) reveals a notable impact of the total data amount on overall computation cost. Conversely, the red lines demonstrate that the resource requirement for Setting 3 (client-binning optimisation) decreased even further compared to Setting 1. This illustrates that a \rev{naive} implementation may \rev{encounter} performance bottlenecks with a large number of data providers, particularly during histogram creation from raw data. However, simple optimisations like client-binning can markedly reduce costs.

Furthermore, we notice that the full-time (time from the beginning until the end of computation) increases more rapidly than the comp-time (time for the main computation only). This disparity, termed as the "setup time", expands with the number of data providers due to the additional time required for encryption agents to prepare and establish connections with computation agents. This discrepancy could stem from factors such as Pod performance, network conditions, operating system constraints, or implementation specifics of MP-SPDZ. By isolating the compute time from setup time, we observe a gradual and slow growth in time with the number of data providers for both Setting 1 (e.g., $374.9$s for 10 data providers versus $379.4$s for 1000 data providers) and Setting 3 (e.g., $101.6$s for 10 data providers versus $106.6$s for 1000 data providers); this also holds for other factors. This slow growth trend in computation time underscores the promising scalability of Libertas, as setup time optimisation can occur independently of MPC computation.

In real-world deployment\rev{s}, the network conditions between computation agents will inevitably impact performance. Task-specific optimisations, like the one demonstrated with client-binning, can substantially enhance performance. Moreover, there is potential for further optimisation by considering the exact computations involved (such as the MWEM algorithm), refining the code, and optimising the MPC protocol to streamline communication rounds and data transmission. We believe this is a sensible expectation in production environment\rev{s}.

The gig worker empowerment scenario exhibited a similar growth trend (Figure \ref{fig:solidmpc_average_wage}). However, a key distinction is that the setup time becomes dominant when the number of data providers is large. This underscores the importance of considering setup costs, particularly for relatively simple computations.

Our evaluation demonstrates that the proposed framework exhibits good scalability, with computation costs scaling linearly in relation to the number of data providers, consistent with our findings in the platform-agnostic benchmark. Moreover, the framework shows potential for significant optimisation based on the specific computational task at hand. Overall, these results provide a promising demonstration of the technical feasibility and scalability of Libertas's implementation with Solid\rev{, the possiblity of the proposed agent services}, as well as \rev{Libertas's} applicability in various computation tasks.

In addition to the empirical results, the case studies also revealed how real-life power dynamics can \rev{facilitate the assignment of trust preferences by} data providers, and also proposed a novel paradigm \rev{for} large-scale collaborative data utilis\rev{ation} while \rev{preserving} input- and output-privacy. That provided confidence for Libertas to create real-world benefits on high-impact use scenarios.
\rev{Relatedly}, we note that privacy assurance and accommodating trust preferences are not only valuable from an ethical perspective but also from the perspective of encouraging adoption and delivering value\rev{, a topic for future studies to explore}.
\del{For example, appropriate privacy assurance and the ability to accommodate different trust preferences can encourage more participation; the utilisation of PDS can lower concerns with data storage and access/usage control, as it is the users' own specified storage.
More participation from data contributors would, in turn, improve the quality of the results by addressing issues like data missingness, delivering more accurate aggregate insights from the collective data. Of course, concrete proofs through (such as) user study will be needed before a definite answer can be stated, as with all technical advancements.}

\section{Conclusions}
User autonomy, decentralisation, privacy-preserving computation, and collaborative computation are all desirable properties for a wide range of tasks and scenarios, but they may seem contradictory at a first glance for one system to achieve.
In this paper, we addressed various challenges associated with building an end-to-end solution for this problem. We discussed the conceptual pitfalls of existing multi-party computation practices, and innovatively proposed an architecture featuring individual-based\rev{,} user-centric trust to overcome such issues in the decentralised context.
We presented a novel architecture called Libertas, which integrates privacy-preserving computation mechanisms like secure multi-party computation (R1 \& R5) with personal data stores \rev{(PDS)} in a modular fashion. Our implementation on Solid maintains compatibility with existing protocols, using the delegated-decentralised model, which provides better scalability as verified by our empirical evaluations. We discussed several features of the proposed solution that will benefit different stakeholders while respecting user autonomy (R2) and providing user-centric trust (R3), \rev{along with} variants for adaptation to different PDS systems. Furthermore, we evaluated our proposed architecture using realistic scenarios, synthetic dataset generation with differential privacy, and gig worker empowerment, demonstrating the wide applicability of our architecture to high-impact collective privacy-preserving computation use cases. Through the case studies, we uncovered how real-life power dynamics can constitute the properties desirable by Libertas, and also how Libertas can achieve large-scale data utilisation while protecting both input and output privacy. The empirical results also verified scalability (R4), while shedding light on possible routes of optimisation when used in production.
Visionarily, our work \rev{offers} a promising direction for empowering users with privacy-preserving and autonomy-respecting collaborative computation, \rev{while} also incentivising \rev{the} adoption of user-centric decentralised technologies like Solid\rev{, by demonstrating realistic use cases and unique benefits. This} is crucial for large-scale\rev{,} privacy-friendly\rev{,} and mutually beneficial collaborative computation and data ecosystems\rev{, and could encourage data contribution with assured privacy and autonomy}.


\subsection*{Limitations and Future Work}
As our work provides a novel technical solution, we also note many interesting future research directions, as well as limitations in our current work.
For example, decentralised systems may face complex governance challenges in ensuring accountability and responsibility \rev{, which} is a very important direction of research. Further, \rev{while} Libertas allows users to \textit{express} and helps them to \textit{utilise} their trusts, \del{but} the current system does not provide additional \rev{support} for users to \textit{\rev{determine}} and \textit{maintain} them. It will be beneficial to explore alternative ways to express, manage and utilise trust of agents, such as different tiers of trust or dynamic preferences based on personnel, MPC computations and protocols, or ``oathbreaker'' recording for strengthened trust\rev{; users may also benefit from a policy checking step using a fine-grained policy language}. Developing dedicated MPC protocols for such a context to utilise its trust dynamics will also be an interesting direction, which may also improve performance. It will also be useful to explore the performance in more broad contexts, such as in WAN setting\rev{s} and production settings; evaluating performance to other modalities of data (such as image data or streaming data) will also be interesting, as the current system focuses on non-real-time non-interactive collective data usage. \rev{Finally}, appropriate user training, real-life collaboration and user studies will benefit the adoption and improvement\del{s} of the architecture\rev{, as well as providing concrete proofs of the societal benefits provided by Libertas-like architectures}.

\begin{acks}
This work is a part of the Ethical Web and Data Infrastructure in the Age of AI (EWADA) project, funded by Oxford Martin School.
In addition, we express our special thanks to Prof. Malcolm Atkinson for his valuable suggestions in improving the presentation of this work.
\end{acks}

\section*{Availability}
The code and specifications of Libertas can be found at \url{https://github.com/OxfordHCC/libertas}. The evaluation logs will be made public in the published version of the paper, or upon request.

\bibliographystyle{ACM-Reference-Format}
\bibliography{bibfile-rui,main}

\appendix

\section*{MWEM Algorithm}
\label{sec:appendix:mwem}

The Multiplicative Weights and Exponential Mechanism (MWEM) algorithm is an iterative algorithm designed to generate a synthetic dataset whose responses to queries closely resemble those of the original dataset \cite{hardt_simple_2012}. MWEM operates by taking the original dataset as input $D$ and a set $Q$ of linear queries.
A linear query, also known as a counting or statistical query, is defined by a function $q$ that maps data records to the interval $[-1,1]$. 
The answer to a linear query on a dataset $D$, denoted by $q(D)$, is the sum $\sum_{x \in D} q(x) \cdot D(x)$. Additionally, two parameters of interest in the algorithm are the epsilon value $\epsilon$ (privacy parameter) and the number of iterations $T$. The algorithm produces a distribution $A$ over $\mathcal{D}$ such that the difference between $q(A)$ and $q(D)$ is small. MWEM repeatedly samples a query for which the difference is still large and updates the weight that $A$  places on each record  $x$. MWEM satisfies $\epsilon$-differential privacy by leveraging the exponential mechanism \cite{mcsherry_mechanism_2007} to sample queries and the Laplace mechanism \cite{dwork_calibrating_2006} to add noise to the query results. The pseudocode of the algorithm is as follows. We implemented this as an MPC circuit without special treatments such as optimisation.

\RestyleAlgo{ruled} 
\begin{algorithm}
\DontPrintSemicolon
  \SetKwInOut{Input}{Input}
    \SetKwInOut{Output}{Output}
  \Input{Dataset $D$ over a universe $\mathcal{D}$, \\ Set of linear queries $Q$, \\ Number of iterations $T$, \\ Privacy parameter $\epsilon > 0$.}
  
  Let $n$ denote $|D|$, the number of records in $D$. \\
  Let $A_0$ denote $n$ times the uniform distribution over $\mathcal{D}$.
 
  \For{$i \in \{1, \ldots, T \}$}
    {
       \textbf{Exponential Mechanism:} sample a query $q_i \in Q$ using the Exponential Mechanism parametrized with epsilon value $\epsilon / 2T $ and the score function: $s_i(D,q) = |q(A_{i-1}) - q(D)|$\\
       \textbf{Laplace Mechanism:} Let measurement $m_i = q_i(D) + \mathsf{Lap}(2T/\epsilon)$\\
       \textbf{Multiplicative Weights:} Let $A_i$ be $n$ times the distribution whose entries satisfy
       $$
       A_{i}(x) \propto A_{i-1}(x) \times \exp(q_{i}(x) \times (m_i - q_i(A_{i-1}))/2n)
       $$
    }
  \Output{$A = A_{T}$ (or $A = \textnormal{avg}_{i<T} A_{i}$)}
\caption{The MWEM algorithm \cite{hardt_simple_2012} \label{MWEMalg}}
\end{algorithm}






\end{document}